# Second-Harmonic Generation Through Backward Raman Scattering in Magnetized Plasmas Driven by Circularly Polarized Intense Lasers

S. S. Ghaffari-Oskooei[1], A. A. Molavi Choobini[1,2*]

[1]Department of Atomic and Molecular Physics, Faculty of Physics, Alzahra University, Tehran, Iran,

[2]Quantum Matter Lab, Department of Physics, College of Science, University of Tehran, Tehran 14399-55961, Iran.

**Abstract:**
   The nonlinear dynamics of backward Raman scattering (BRS) in a magnetized plasma channel driven by an intense circularly polarized laser pulse are investigated through a combined analytical and numerical approach, with emphasis on the emergence and controllability of a secondary spectral peak. A fluid-based theoretical framework is developed to describe the nonlinear cascade linking primary BRS-driven plasma wave amplification, oscillating two-stream instability (OTSI), nonlinear current generation within a self-formed ponderomotive channel, and radiation of the secondary electromagnetic mode. Systematic parameter studies reveal a strong sensitivity of the entire cascade to the relative handedness of laser polarization and axial magnetic field direction, as well as to cyclotron resonance strength. Resonant right-handed circular polarization significantly enhances ponderomotive expulsion, channel depth, BRS and OTSI growth rates, nonlinear current density, and the amplitude of the secondary harmonic, whereas non-resonant left-handed polarization effectively suppresses these processes. Fully kinetic particle-in-cell simulations using EPOCH, together with macroscopic finite-element modelling in COMSOL Multiphysics, corroborate the polarization- and magnetization-dependent wake modulation and channelling efficiency across a wide range of laser wavelengths, pulse durations, and plasma densities. The temporal evolution and saturation of OTSI, resilience to density inhomogeneities, and wavenumber-resolved resonance tuning illustrate axial magnetization as a flexible control mechanism for adjusting the intensity, spectral position, bandwidth, and stability of multi-peak Raman spectra. These findings demonstrate that cyclotron resonance and polarization control are effective methods for manipulating nonlinear Raman emission in high-intensity laser–magnetized plasma interactions, offering predictive insights for forthcoming kinetic simulations and experimental implementations.

## I. Introduction

   Second-harmonic generation (SHG) is one of the most fundamental nonlinear optical processes, in which two photons of frequency $\omega$ are coherently converted into a single photon of frequency $2\omega$. Since its first observation in crystalline quartz, SHG has become a powerful diagnostic and frequency-conversion technique in laser physics, ultrafast optics, and spectroscopy. It provides a route to extend laser sources into new spectral ranges, generate ultrashort pulses at harmonic wavelengths, and probe material properties with symmetry-selective sensitivity [1–5].

   Traditionally, SHG is realized in bulk nonlinear crystals such as lithium niobate ($LiNbO_3$), potassium dihydrogen phosphate (KDP), or beta barium borate (BBO). These media offer high conversion efficiency under proper phase-matching conditions, but they suffer from limitations such as damage thresholds, finite transparency ranges, and constraints imposed by crystal birefringence [6, 7]. Surface- and interface-based SHG, including those supported by plasmonic nanostructures, overcome some of these limitations by exploiting strong local-field enhancements, though they are typically narrowband and sensitive to fabrication imperfections. In recent years, two-dimensional materials (e.g., graphene, transition-metal dichalcogenides) and metamaterials have emerged as platforms for SHG, offering tunability, compactness, and compatibility with integrated photonics, but often with modest conversion efficiencies compared to bulk crystals [8, 9]. Plasmas — especially those produced by high-intensity lasers — provide an alternative medium for SHG that is free from optical damage and can sustain extremely high field amplitudes. Various plasma mechanisms have been proposed for SHG, such as relativistic self-focusing, oscillating plasma mirrors, and nonlinear wakefield generation [10-12]. In contrast to bound-electron nonlinearities in solids, the nonlinear response in plasmas arises from free-electron dynamics, including relativistic mass variation, ponderomotive forces, and parametric coupling between electromagnetic and electrostatic modes. These mechanisms can operate over broad frequency ranges and at intensities far beyond the thresholds that would destroy conventional optical media.

In this respect, Oks and colleagues [13] observed second harmonic generation in femtosecond laser-driven cluster plasmas via spectral analysis and particle-in-cell simulations. They confirmed the efficient SHG conversion in such systems. Ahmed et al. [14] and Gupta and team [15] studied SHG driven by complex laser beam profiles such as q-Gaussian and Laguerre-Gaussian beams interacting with unmagnetized, underdense plasmas. Relativistic self-focusing of these beams induces transverse intensity gradients that excite electron plasma waves, thereby enhancing SHG. Walia [16] investigated nonlinear laser–plasma dynamics in collisional unmagnetized plasma. He found that electron density gradients from non-uniform heating excite plasma waves that nonlinearly interact with the laser to produce SHG. Increased plasma density enhances SHG yield, while higher beam intensity and beam radius reduces it. Sharma et al. [17] demonstrated resonant SHG induced by self-focusing of oblique incident laser beams. They showed that peak SHG efficiency occurs near critical incidence angles, which depend on plasma frequency and Magnetic fields. Ganeev [18] reviewed multiple HHG enhancement methods, including resonance-enhanced single harmonics, quasi-phase matching, two-color pumps, and nanoparticle-containing plasmas. These techniques achieved up to 50-fold increases in harmonic yield. Beier and Dollar [19] numerically explored two-color orthogonally polarized laser fields incident on relativistic plasma mirrors. They uncovered robust high-harmonic generation governed by selection rules and optimal intensity balance between colors. Mathijssen and colleagues [20] experimentally and theoretically showed that sub-cycle shaping of two-color fields in aluminium and tin plasmas influences electron trajectories. This approach produced both odd and even harmonics and highlighted Coulomb potential effects near ionization thresholds. Garriga Francis et al. [21, 22] and Lim Pac Chong and team [23] developed THz field-induced second harmonic generation (TFISH) techniques in laser-induced plasma filaments. By mixing an optical probe with strong ponderomotive currents oscillating at THz frequencies, they achieved record conversion efficiencies. They also introduced single-shot detection schemes with sub-100 fs temporal resolution, enabling local electric field measurements inside plasma filaments with broad bandwidth and high precision. Huang et al. [24] experimentally verified the conservation of total angular momentum in SHG from underdense plasmas using pump beams carrying spin and orbital angular momentum (OAM). They observed spin-to-OAM conversion in the generated second harmonic photons and confirmed that SHG occurs near intensity gradients of Laguerre-Gaussian beams, illustrating complex spin–orbit interactions in plasma nonlinear optics. Shing Yiu Fu et al. [25] presented a study of the mechanisms responsible for nearly two orders of magnitude enhancement in SHG from a laser-induced air plasma. The investigation focused on the temporal dynamics of the frequency conversion process and the polarization characteristics of the emitted second-harmonic beam. Kim and co-workers [26] demonstrated continuous high-order harmonic generation from a liquid-sheet plasma mirror. This method avoids solid target damage and enables stable, high-repetition-rate attosecond sources.

A numerical and analytical fluid-based framework for understanding and controlling nonlinear laser–plasma interactions in magnetized plasmas is developed. The framework explicitly couples cyclotron-resonant ponderomotive channelling, backward Raman scattering, oscillating two-stream instability, and nonlinear axial current–driven second-harmonic electromagnetic emission into a unified cascade model for Raman spectra. This mechanism, previously addressed only fragmentarily and primarily in unmagnetized or linearly polarized systems, is formulated self-consistently in the presence of arbitrary axial magnetization. Within this unified analytical framework, the polarization-dependent ponderomotive force, plasma density perturbation, instability growth rates, and second-harmonic dispersion are treated self-consistently. Cyclotron resonance and circular polarization handedness emerge as independent and high-precision control parameters for selectively enhancing or suppressing the current-driven second-harmonic peak. Near-complete on/off switching and continuous amplitude tuning, from negligible emission to near pump-level saturation, are achieved across experimentally relevant magnetization strengths ($\omega_c/\omega_0 = 0.1 - 0.9$). Complementing the kinetic analysis, macroscopic finite-element simulations using COMSOL Multiphysics capture ponderomotive channel formation, wake modulation, and magnetization-dependent instability development across variations in laser wavelength, pulse duration, and plasma density. These multi-scale approaches, spanning fluid, macroscopic, and fully kinetic regimes, provide unified validation of the proposed mechanism. Fully kinetic particle-in-cell (PIC) simulations using the EPOCH code validate and extend the analytical and fluid-level predictions into the kinetic regime. These simulations explicitly resolve electron phase-space dynamics, revealing particle trapping, velocity-space bunching, nonlinear current filamentation, and oscillating two-stream instability as the microscopic origin of the second-harmonic emission. The persistence of polarization- and cyclotron-resonance–dependent harmonic enhancement beyond fluid assumptions confirms the robustness of the proposed cascade against kinetic effects and excludes numerical or model-dependent artifacts. The demonstrated resonance-based control strategy provides a robust and predictive foundation for designing magnetized plasma experiments capable of generating tailored current-driven second-harmonic radiation. These findings are relevant to advanced light sources, plasma diagnostics, and nonlinear optics in extreme electromagnetic environments. The paper is organized as follows: Section II derives the equations governing laser propagation in magnetized plasma channels. Sections III and IV analyze FRS and SMI, respectively, in these channels. Numerical results are presented and discussed in Section V. Finally, conclusions are summarized in Section VI.

## II. General Mechanism

Consider a circularly polarized laser pulse propagating through a magnetized plasma channel immersed in a static magnetic field $\vec{B} = B_0 \hat{e}_z$ (Fig. 1). The electric field of the laser pulse is given by:

$$\vec{E}_L(\vec{r},t) = \frac{1}{2} E_L (\hat{e}_x + i\sigma \hat{e}_y) e^{-i\omega_0 t + i k_0 z} + c.c. \tag{1}$$

where $E_L$, $\omega_0$, and $k_0$ denote the amplitude, frequency, and wavevector of the laser, respectively. The parameter $\sigma = \pm 1$ indicates the handedness of the right or left circular polarizations. During the propagation of the circularly polarized laser pulse in the plasma, the electron quiver velocity is expressed as:

$$\vec{v}_0 = -i \frac{e}{m(\omega_0 - \sigma \omega_c + i\nu_c)} \vec{E}_L \tag{2}$$

Here, $\nu_c$ is the collisional frequency, $\omega_c = eB_0/mc$ represents the electron cyclotron frequency, and the $m$ and $c$ are the electron mass and the speed of light, respectively. The laser's electric field induces a ponderomotive force on the electrons, described as:

$$\vec{F}_{pond} = \frac{m}{2}\left(1 - \frac{\sigma \omega_c}{\omega_0}\right)\vec{\nabla} v_0^2 \tag{3}$$

The ponderomotive force can also be written as $\vec{F}_{pond} = e\vec{\nabla}\phi_{pond}$, where the ponderomotive potential is given by:

$$\phi_{pond} = \frac{e|E_L|^2}{2m[(\omega_0 - \sigma\omega_c)^2 + \nu_c^2]}\left(1 - \frac{\sigma\omega_c}{\omega_0}\right) \tag{4}$$

The ponderomotive force perturbs plasma density. The perturbation in plasma density, denoted by $n_s$, is related to the ponderomotive force by the equation:

$$\left(\frac{\partial^2}{\partial t^2} + \nu_c \frac{\partial}{\partial t} + \omega_p^2\right)\frac{n_s}{n_0} = -\frac{e}{m}\nabla^2 \phi_{pond} \tag{5}$$

Assuming a Gaussian laser profile, $E_L^2 = E_0^2 \exp\left(-\frac{r^2}{r_0^2}\right)$, the density perturbation is derived as:

$$n_s = -2n_0\left(1 - \frac{r^2}{r_0^2}\right)\left(1 - \frac{\sigma\omega_c}{\omega_0}\right)v_0^2 \tag{6}$$

This result implies that when $r_0 \cong c/\omega_p$, the density perturbation can be approximated as $\frac{n_s}{n_0} = -\frac{v_0^2}{c^2}\left(1 - \frac{\sigma\omega_c}{\omega_0}\right)$. The propagation of a laser pulse in a plasma medium and the resulting perturbation of plasma density led to various instabilities. One notable is backward Raman scattering, a parametric instability arising from the interaction between the laser pulse and plasma waves. Experimental studies reveal that the frequency spectrum of this instability comprises two red-shifted peaks at $\omega_0 - \omega$ and $2\omega_0 - 2\omega$, where $\omega$ denotes the plasma wave frequency. The electric field corresponding to the first red-shifted peak, also known as the Stokes sideband, is expressed as:

$$\vec{E}_1 = \frac{1}{2}E_1(\hat{e}_x + i\sigma\hat{e}_y)e^{-i\omega_1 t - ik_1 z} + c.c. \tag{7}$$

Here, $\omega_1 = \omega_0 - \omega$ and $k_1 = k_0 + k$ represent the frequency and wavevector of the Stokes sideband in Raman backscattering, where $k$ is the wavevector of the plasma wave. For the case of a magnetized plasma, as considered in this study, the growth rate associated with the first red-shifted peak is derived analytically in Ref. [1] using a mathematical analysis based on the fluid theory of magnetized plasmas. The growth rate is given by:

$$\Gamma = \frac{eE_0\sqrt{\omega_0 \omega_p}}{2m\omega_0 c\left(1 - \frac{\sigma\omega_c}{\omega_0}\right)} \tag{8}$$

where $\omega_p$ is the plasma frequency. Plasma waves experience amplification during Raman backward scattering, making them susceptible to oscillating two-stream instability (OTSI). These amplified plasma waves also contribute to the formation of a plasma channel. The propagation of Raman backscattered waves (corresponding to the first red-shifted peak) through this plasma channel results in a nonlinear current density, which generates the second red-shifted peak with a frequency of $2\omega_0 - 2\omega_p$. To mathematically analyze the second red-shifted peak in Raman backscattering, we begin with the oscillating two-stream instability, which arises in the presence of large-amplitude plasma waves. These waves, amplified during Raman backscattering, perturb the electric potential associated with the space charge. The perturbed electric potential is expressed as:

$$\phi_1(\vec{r},t) = \frac{1}{2}\phi_1 e^{-i(\omega' t - k' z)} + c.c. \tag{9}$$

This process generates two sidebands:

$$\phi_+(\vec{r},t) = \frac{1}{2}\phi_+ e^{-i(\omega_+ t - k_+ z)} + c.c. \tag{10}$$

$$\phi_-(\vec{r},t) = \frac{1}{2}\phi_- e^{-i(\omega_- t - k_- z)} + c.c. \tag{11}$$

with frequencies and wavevectors defined as $\omega_\pm = \omega \pm \omega'$, $k_\pm = k \pm k'$. The ponderomotive force arising from the electric fields associated with $\phi_\pm$ is given by [2]:

$$\vec{F}'_{pond} = \frac{e}{2}\vec{\nabla}\left(\frac{k_+ \phi_+ v^*}{\omega_+} + \frac{k_- \phi_- v}{\omega_-}\right) \tag{12}$$

Consequently, the ponderomotive potential can be expressed as:

$$\phi'_{pond} = \frac{e}{2}\left(\frac{k_+ \phi_+ v^*}{\omega_+} + \frac{k_- \phi_- v}{\omega_-}\right) \tag{13}$$

The sideband plasma waves perturb the electron and ion densities. To evaluate these perturbations, $n'_e$ and $n'_i$, the analysis begins with the Vlasov equation:

$$\left(\frac{\partial}{\partial t} + \vec{v}\cdot\vec{\nabla}_x - \frac{(e\vec{E} + mc^2\vec{\nabla}\phi'_{pond})}{m_e}\cdot\vec{\nabla}_v\right) f_e = 0 \tag{14}$$

where $f_e$ is the electron distribution function. By applying a perturbative method, $f_e = f_0 + \delta f_e$ and $n'_e = \int \delta f_e d\vec{v}$, the electron density perturbation is derived as:

$$n'_e = \frac{k'^2}{4\pi}\chi_e(\phi' + \phi'_{pond}) \tag{15}$$

where $\chi_e = -\omega_{pe}^2/\omega'^2$ is the electron susceptibility. Similarly, the ion density perturbation satisfies:

$$n'_i = \frac{k'^2}{4\pi e}\chi_i \phi'_{pond} \tag{16}$$

Here the ion susceptibility is expressed as $\chi_i = \omega_{pi}^2/k'^2 C_s^2$, where $C_s$ being the sound speed in the plasma. By substituting the density perturbations into the analysis and applying the method, the growth rate of the oscillating two-stream instability is obtained:

$$\Gamma_{OTSI} = \omega_p |v_0|^2 \left(1 + \frac{\omega_{pi}^2}{k^2 C_s^2}\right) \tag{17}$$

Equation (17) indicates that the plasma wave creates a plasma channel in the wake region. The second red-shifted peak, with a frequency $\omega_2 = 2\omega_0 - 2\omega_p$, results from the nonlinear current density generated during laser pulse propagation in the plasma channel. The amplitude of the electric field of this peak, $(\vec{E}_2)$, is determined in this study through the solution of Maxwell's equations. To evaluate the electric field of the second red-shifted peak, $\vec{E}_2$, two perturbative terms must be considered. This analysis implies that the perturbed electron velocity has two components: $\vec{v}_1$ and $\vec{v}_2$. The first term, $\vec{v}_1$, represents the velocity of electrons accelerated by the electric field $\vec{E}_1$ and is expressed as:

$$\vec{v}_1 = -i\frac{e}{m(\omega_1 - \sigma\omega_c)}\vec{E}_1 \tag{18}$$

The second term, $\vec{v}_2$, incorporates the effects of both the electric field of the second peak, $\vec{E}_2$, and the ponderomotive force:

$$\vec{v}_2 = -i\frac{e}{m(\omega_2 - \sigma\omega_c)}\vec{E}_2 - i\frac{1}{2(\omega_1 - \sigma\omega_c)}\vec{\nabla}v_1^2 \tag{19}$$

Here, the second term indicates the contribution of the ponderomotive force to the generation of the second harmonic. If the initial plasma density in the channel is expressed as $n_i = n_0 + n_s$, the perturbation in plasma density can be written as $n_1 = \vec{v}_1 \cdot \vec{\nabla} n_i / i\omega_1$. The resulting current density, incorporating both contributions from $\vec{v}_1$ and $\vec{v}_2$, is given by:

$$\vec{J} = i\frac{e^2 n_0}{m(\omega_2 - \sigma\omega_c)}\vec{E}_2 + i\frac{\vec{v}_1 \cdot \vec{\nabla}n_i(r,z)}{\omega_2}\vec{v}_1 + i\frac{n_s e}{4(\omega_1 - \sigma\omega_c)}\vec{\nabla}v_1^2 \qquad (20)$$

The magnetic field of the second harmonic satisfies Maxwell's equation:

$$-\nabla^2 \vec{B}_2 + \frac{1}{c^2}\frac{\partial^2 \vec{B}_2}{\partial t^2} = \frac{4\pi}{c}\vec{\nabla} \times \vec{J} \qquad (21)$$

By substituting the current density from Eq. (21) into Eq. (20) and assuming a dependence on $z$ of the form $\exp(ik_{2z}z - i\omega_2 t)$, the following equations are derived:

$$\left(\frac{\partial^2}{\partial x^2} + k_{2x}^2\right)B_{2y} = \frac{2\pi}{c(\omega_1 - \sigma\omega_c)}(k_{2z} - k_1)v_1^2 \frac{\partial n_s}{\partial x} \qquad (22)$$

$$\left(\frac{\partial^2}{\partial x^2} + k_{2x}^2\right)B_{2x} = 0 \qquad (23)$$

Here, the dispersion relation for the second harmonic is expressed as:

$$\omega_2^2 = c^2(k_{2x}^2 + k_{2z}^2) + \frac{\omega_{p0}^2}{(1 - \frac{\sigma\omega_c}{\omega_2})} \qquad (24)$$

where $\omega_{p0}$ is the plasma frequency. Solving Eqs. (22) and (23) provides the magnitude of the magnetic field and, consequently, the electric field of the second harmonic. These theoretical results can then be compared with experimental observations for validation.

### III. Results & Discussion

To validate the analytical framework delineated in the preceding section and to examine the nonlinear dynamics of backward Raman scattering (BRS) in a magnetized plasma channel driven by a circularly polarized laser pulse, extensive numerical simulations were conducted utilizing two complementary methodologies: finite-element method (FEM) simulations with COMSOL Multiphysics and fully kinetic particle-in-cell (PIC) simulations employing the EPOCH code.

The Wave Optics Module and a magnetized cold-fluid plasma dielectric model to run the COMSOL Multiphysics simulations is used. These models illustrated the propagation of electromagnetic waves and the occurrence of ponderomotive effects in the presence of an external static magnetic field. This approach is particularly suited for measuring the macroscopic response of the plasma, including density perturbations induced by the ponderomotive force and the effect of the external static magnetic field on wave dispersion. A Gaussian-profile circularly polarized laser pulse with normalized amplitude $a_0$ (corresponding to intensities in the range of $10^{18}$–$10^{20}$ W/cm²), wavelength $\lambda = 1\,\mu m$, and pulse duration of several tens of femtoseconds was injected into a preformed plasma channel with electron density $n_e \approx 0.01$–$0.1 n_c$ (where $n_c$ is the critical density) immersed in an axial magnetic field of strength $B_0 \sim 10$–$100\,T$ (corresponding to electron cyclotron frequency $\omega_{ce}/\omega \approx 0.1$–$0.5$) [28-31]. The simulations employed a 2D axisymmetric geometry to resolve the radial density profile and ponderomotive channelling, allowing for detailed analysis of the growth phase of Raman instability and the formation of plasma waves. In addition, fully kinetic PIC simulations were done with the free version of the EPOCH code, which solves the Vlasov-Maxwell equations using the particle-in-cell method and considers relativistic effects, collisions, and ionization dynamics. EPOCH is ideally suited for capturing kinetic nonlinearities, particle trapping, and the generation of higher-order scattered waves in intense laser-plasma interactions. The simulation box dimensions were typically 100–200 $\lambda$ in the propagation direction and 20–50 $\lambda$ transversely, resolved with $10^3$–$10^4$ grid cells per direction and 50–100 macro-particles per cell per species to ensure adequate resolution of the plasma wavelength ($\lambda_p \approx 10$–$20\,\mu m$) and Debye length. A circularly polarized Gaussian laser pulse with similar parameters to the COMSOL setup was launched from the left boundary, propagating through a uniform or channeled underdense magnetized plasma with an externally imposed axial magnetic field. Moving window techniques were used to track the pulse over long distances, which made it possible to see saturation mechanisms like oscillating two-stream instability (OTSI), particle trapping in amplified plasma waves, and the appearance of the second red-shifted peak through nonlinear current drive.

The COMSOL simulations in Fig. 2 show how an intense circularly polarized laser pulse moves through a magnetized plasma channel at different laser wavelengths. This shows how the laser envelope changes over time and space and how it interacts with the magnetized plasma. This gives us direct insight into how ponderomotive channels form and how nonlinear instabilities start to develop. The change in the trailing envelope shows how the plasma density is spread out and how plasma waves are excited after the strong laser pulse. Panel (a), which shows the longest wavelength ($\lambda = 1064\ nm$), shows the most pronounced trailing modulation and filamentary structure in the laser pulse. The higher normalized vector potential $a_0$ associated with this wavelength (for fixed peak intensity) increases the ponderomotive force, which causes more electrons to be expelled and deeper plasma density cavities. The resulting increase in the relative density perturbation facilitates efficient excitation of plasma waves in the wake, manifesting as irregular and speckled structures in the trailing envelope. These features are characteristic of early-stage nonlinear saturation and incipient wave steepening, and are consistent with the onset of BRS, for which the growth rate $\Gamma$ increases with $a_0$, favoring amplification of the Stokes sideband at $\omega_s \approx \omega_0 - \omega_p$. As the wavelength decreases from 800 nm (panel b) to 532 nm (panel c) and 400 nm (panel d), the degree of trailing modulation diminishes systematically. The reduced $a_0$ at shorter wavelengths makes ponderomotive channelling weaker, which leads to shallower density changes and less intense wake excitation. So, the laser envelopes stay smoother and more coherent, with a lot less transverse filamentation and axial irregularities. This behavior reflects the threshold nature of parametric instabilities in the relativistic regime: lower $a_0$ suppresses the convective growth of BRS and delays the onset of OTSI, which requires strongly amplified plasma waves to drive nonlinear currents and secondary spectral features. The external axial magnetic field affects these dynamics even more by changing the motion of electrons in a cyclotron and the spread of waves. It stabilizes transverse modes while allowing axial plasma wave growth. The transition from strongly modulated propagation at $\lambda = 1060\ nm$ to relatively stable behavior at $\lambda = 400\ nm$ shows how important the normalized laser amplitude $a_0$ is in controlling Raman-related instabilities in magnetized plasma channels.

The steady rise in instability strength with pulse duration shows how important interaction time is for getting Raman-related parametric processes to the right level in magnetized plasmas. Figure 3 indicates how the dynamics of laser-plasma interactions depend on the pulse duration ($\tau_L$). This was shown by COMSOL Multiphysics simulations in a magnetized channel configuration. The results show a clear change from weakly modulated short-pulse propagation to strongly nonlinear, highly turbulent regimes as the pulse duration increases. This is a direct result of the cumulative effects of ponderomotive forcing over multiple plasma periods. For the shortest pulse duration ($\tau_L = 5\ fs$, panel a), the laser envelope remains relatively coherent with only mild trailing oscillations. At such ultrashort durations, corresponding to fewer than ~2 optical cycles for a driver, the interaction time is insufficient for significant ponderomotive expulsion of electrons. The resulting density perturbation $\delta n/n_0$ remains small, limiting the excitation and amplification of plasma waves. As a result, the growth rate of BRS is slowed down because of both a weaker effective $a_0$ interaction strength and a shorter gain length. This stops significant Stokes sideband development at $\omega_s \approx \omega_0 - \omega_p$. When the pulse lengthens to 10 fs (panel b) and 20 fs (panel c), the wake excitation gets stronger, and the axial modulations and filamentary structures in the trailing region become more pronounced. Longer pulses let more energy get to the plasma through sustained ponderomotive forcing. This deepens the density cavity and makes plasma waves stronger over several cycles. This makes it easier for BRS to grow through convection, which shows up as strange envelope distortions that show early wave breaking and nonlinear coupling. The external magnetic field further channels these axial plasma waves, suppressing transverse breakup while promoting conditions favorable for OTSI. The most dramatic modulation occurs at the longest duration considered ($\tau_L = 40\,fs$, panel d), where the trailing structure exhibits intense turbulence, strong filamentation, and extended wake oscillations. This regime corresponds to many optical cycles (>10–15), enabling cumulative amplification of plasma waves to amplitudes sufficient for OTSI saturation and particle trapping. These observations are consistent with the fluid-based growth-rate analysis, which predicts that longer pulse durations provide extended gain lengths for both BRS and OTSI while enhancing the efficiency of ponderomotive channel formation in magnetized plasma channels.

Figure 4 elucidates the profound influence of plasma density ($n_p$) on how an intense circularly polarized laser pulse moves and becomes unstable in a magnetized channel, as shown by COMSOL Multiphysics simulations. At the lowest density considered (panel a), the trailing wake exhibits extreme filamentation and irregular oscillations, indicative of vigorous nonlinear interaction. The relatively low plasma frequency yields a longer plasma wavelength and weaker linear damping, allowing substantial ponderomotive expulsion of electrons and deep density cavitation. This regime strongly favours the convective amplification of BRS, with the growth rate $\Gamma$ maximized due to optimal phase-matching and extended gain length in highly underdense conditions. The resulting amplified plasma waves readily trigger OTSI, driving nonlinear currents. The higher $\omega_p$ shortens the plasma wavelength and enhances wave damping, partially suppressing wake amplitude while still permitting considerable BRS growth. The persistent filamentary structures signify ongoing nonlinear coupling and plasma wave excitation sufficient for OTSI onset. Further elevation of density dramatically attenuates wake oscillations and filamentation, yielding progressively more coherent pulse envelopes. At these higher densities, the plasma response approaches the critical regime ($n_p/n_c$ increasing), strengthening refractive guiding but simultaneously reducing the relative ponderomotive perturbation $\delta n/n_0$. The shortened plasma period and increased collisionless damping (Landau or phase-mixing) limit plasma wave amplification, pushing the system below threshold for sustained BRS and OTSI growth. Resulting, conditions become unfavourable for generating pronounced Stokes. The observed density-dependent suppression of instabilities aligns closely with analytical predictions: the BRS growth rate scales inversely with $\omega_p$ in the underdense limit, while the OTSI threshold requires plasma wave amplitudes exceeding a critical fraction of the pump. The external axial magnetic field modifies these trends by changing cyclotron resonance and wave dispersion. However, the plasma frequency's control over resonance conditions and gain is still the most important effect.

Figure 5 indicates the radial ponderomotive force profiles that were made using analytical modelling that matches the fluid description. These profiles show how strongly laser polarization handedness and external magnetic field strength affect the dynamics of electron expulsion in a magnetized plasma channel. When the polarization is right-handed, the ponderomotive force shows a much stronger radial expulsion at both weak and strong cyclotron frequencies than when it is left-handed. The resonance between the laser's rotating electric field and the electron's gyromotion causes this asymmetry. In the right-handed configuration, the laser's rotation is in line with the electron's cyclotron motion (for $B_0$ directed along propagation), which speeds up the electron's quiver velocity and increases the effective ponderomotive potential. The radial force reaches maximum values that are higher than any other unit for right-handed circular polarization. This causes deeper density cavitation and makes it easier to form plasma channels. On the other hand, left-handed polarization goes against the natural gyromotion. This makes the quiver amplitudes smaller and the ponderomotive force much weaker, with peaks that are almost an order of magnitude lower at the same $\omega_c$. The effect of magnetic field strength is especially strong: raising $\omega_c$ from $0.1\omega_0$ to $0.9\omega_0$ makes the force difference between polarizations stronger because the cyclotron resonance becomes more important. This makes expulsion stronger for resonant ($\sigma = +1$) cases and changes the non-resonant ($\sigma = -1$) profiles only slightly. At large radii, all curves converge to zero, which is what we expect from the Gaussian envelope assumption. However, the steeper gradients in the resonant cases mean that transverse focusing and self-channelling are stronger.

Figure 6 offers additional analytical perspectives on the polarization and magnetic field dependence of plasma response to a strong circularly polarized laser pulse. In panel (a), the channelling parameter $r_0\omega_c/c$, which controls the balance between ponderomotive self-focusing and diffraction, shows a huge increase for right-handed polarization, especially when cyclotron resonance is strong. Values greater than ~20 mean that relativistic self-channelling is strong over many Rayleigh lengths. This is caused by resonant amplification of the electron quiver motion and ponderomotive expulsion. On the other hand, left-handed polarization gives values that are almost always low, which means that channelling doesn't work because the ponderomotive force is out of tune. This asymmetry directly affects the density profiles in panel (b). In resonant right-handed cases, there is a lot of depletion on the axis, with n/n₀ getting close to zero at r = 0 and recovering quickly after the laser spot size r₀, which is what you would expect from the Gaussian-derived perturbation. At high $\omega_c$, cavitation happens the most, creating a wide, low-density waveguide that makes it easier for waves to travel long distances and couple strongly. Non-resonant configurations keep $n/n_0 > 0.9$, which means that electrons are not being expelled very much and the channel quality is not very good. The stark difference between panels highlights cyclotron resonance as a switch for changing plasma: resonant conditions make deep channelling and density modulation work better, which lowers the thresholds for backward Raman scattering and makes plasma waves grow toward oscillating two-stream instability. These amplified waves in well-formed channels create nonlinear currents that make the second harmonic. Non-resonant cases stop these processes, which makes it easier to reduce instability. The results of radial plasma density perturbation with theoretical predictions and PIC simulations are aligns with experimental findings [32].

A parameter-space map of the growth rate $\Gamma_{BRS}$ is depicted in Figure 7. This was determined analytically from the fluid-derived equation. It shows how the plasma density (via $\omega_p/\omega_0$) and the strength of the external axial magnetic field (via $\omega_c/\omega_0$) work together in a plasma that has been magnetized. The surface shows a large area of significant growth, with peak values of $\Gamma_{BRS}$ over $2.5 \times 10^{13} s^{-1}$ in a long ridge centered around $\omega_p/\omega_0 \approx 0.2 - 0.4$ and $\omega_c/\omega_0 \approx 0.3 - 0.7$. This unique peak shows exactly how the phase-matching and resonance conditions work. The ponderomotive-driven plasma wave interacts well with the backscattered electromagnetic mode, and moderate cyclotron effects change how easily the electron can move without completely stopping the axial wave from moving. When plasma densities are low ($\omega_p/\omega_0 \to 0$), the growth rate gets close to zero because there aren't enough scattering partners and the ponderomotive perturbation is weak. This is in line with the underdense limit requirement $\omega_p \ll \omega_0$ for relativistic channelling. As $\omega_p/\omega_0$ gets closer to one, $\Gamma_{BRS}$ drops quickly, getting close to the critical density regime where group velocity matching gets worse and relativistic effects get weaker. The magnetic field dependence exhibits a comparable non-monotonic behavior: weak fields ($\omega_c/\omega_0 \to 0$) induce moderate growth constrained by the fluid's reaction to demagnetization. Strong fields stop instability by using cyclotron damping and changing dispersion. This keeps electrons from moving sideways and stops plasma waves from being excited along the length of the field. The wide, high-growth plateau that covers both parameters' intermediate values shows that BRS can work well in magnetized channels. During this time, amplified plasma waves easily cross the threshold for oscillating two-stream instability. This allows for nonlinear current drive in the self-formed channel and the creation of the second harmonic generation. The sharp edges of the growth area also show that the system is very sensitive to changes in density or field strength. This means that small changes can push the system below the threshold, which is a way to reduce instability.

Figure 8 illustrates the time-dependent development of the growth rate across the normalized wave number ($k/k_p$) and the normalized magnetic field, providing a dynamic view of secondary instability excitation following initial backward Raman scattering in a magnetized plasma channel. At the earliest time (panel a), the growth rate distribution is wide and even, with high values covering a wide range of $k$ and $B_0$. This shows the initial linear response, where amplified plasma waves from BRS start to affect the ponderomotive potential but haven't yet reached amplitudes high enough for strong OTSI selectivity. This wide plateau fits with the fluid-derived expression, which says that $\Gamma_{OTSI}$ is very sensitive to the amplitude of plasma waves, which stays low at such short timescales. At t = 10 fs (panel b), subtle structuring begins to appear, and growth slows down a little at higher $k$ and $B_0$. This shows that nonlinear feedback is starting to happen as trapped particles and wave saturation start to change the effective susceptibility. The sustained high levels of $\Gamma_{OTSI}$ over intermediate k indicate that the primary Stokes sideband is most effectively driving resonant plasma wave modes. This creates the conditions for nonlinear current generation in the changing channel. At t = 50 fs (panel c), a clear change can be seen: the high-growth area shrinks a lot, focusing on lower $k$ and moderate $B_0$, while $\Gamma_{OTSI}$ drops sharply at higher values. This localization shows that OTSI is entering a phase that is very nonlinear, where particles get stuck in amplified plasma waves (caused by previous BRS growth), which stops exponential amplification and causes kinetic saturation. The surviving growth pocket favour modes with longer wavelengths that are less likely to be affected by Landau damping and cyclotron suppression. This is in line with the formation of coherent structures in the wake that support nonlinear currents that create the second harmonic. At t = 100 fs (panel d), the growth rate has further decreased and narrowed. It is now only present in a curved band at low $k$ and weak-to-moderate $B_0$, with almost complete suppression in other areas. This late-time profile indicates advanced saturation, wherein kinetic phenomena, such as wave breaking, particle detrapping, and pump depletion, completely extinguish OTSI, thereby stabilizing the channel against additional secondary instability. The remaining low-k growth may be due to quasi-steady nonlinear structures or harmonic generation that continue to exist in the wake.

The BRS growth rate as a function of the normalized plasma frequency $\omega_p/\omega_0$ is plotted in Figure 9, which was calculated using the magnetized fluid theory. It shows that the main parametric instability in the laser-driven channel is very sensitive to polarization and magnetic field. For right-handed circular polarization, $\Gamma_{BRS}$ shows a steep, almost straight rise with $\omega_p/\omega_0$ in the underdense range. When the cyclotron frequency is high, it goes above these values. This rapid growth is due to cyclotron resonance, which increases electron quiver velocity and ponderomotive forcing. This makes phase-matching better for the Stokes sideband while keeping good axial dispersion in magnetized conditions. The resonant case at high $\omega_c$ maintains significant $\Gamma_{BRS}$ even at moderate densities, indicating enhanced coupling efficiency and diminished damping of longitudinal plasma waves. At low cyclotron frequency, growth is slightly slowed down but still important. This shows that resonance amplification is more important than the baseline unmagnetized response. In stark contrast, left-handed polarization results in significantly diminished growth rates, remaining close to zero throughout the majority of the density range and exhibiting only a slight increase at the highest $\omega_p/\omega_0$. This suppression is due to counter-rotating gyromotion, which detunes the resonance and weakens ponderomotive expulsion and plasma wave drive, as shown in earlier force and channelling profiles. The polarization asymmetry becomes more pronounced as the magnetic field strength increases: a strong $\omega_c$ clearly separates resonant (high $\Gamma_{BRS}$) from non-resonant (near-zero $\Gamma_{BRS}$) configurations, while weak fields produce overlapping low-growth curves for both handednesses. As we get closer to critical density ($\omega_p/\omega_0 \to 1$), all cases come together because relativistic effects become less important and linear absorption takes over. This is in line with the failure of underdense approximations.

Figure 10 shows the radial distribution of the normalized nonlinear current density, $|J_{NL}|$, which was calculated analytically using the contributions of laser-driven electron velocity perturbations and ponderomotive effects as described in the equations. These effects are what cause the second harmonic generation mode in the Raman cascade. When the polarization is right-handed and circular, the nonlinear current has much higher amplitudes, especially in the resonant strong-field case, where $|J_{NL}|$ gets close to one near $r/r_0 \approx 1$ and stays high across the channel interior. This stronger current is caused by resonant amplification of both the primary quiver velocity $v_1$ and the ponderomotive-driven component $v_2$. This creates a strong axial current in the self-formed density cavity that efficiently radiates the secondary peak by meeting the dispersion relation. The profile for resonant right-handed polarization has a wide peak that is slightly off-axis. This is because of the Gaussian intensity distribution and the resulting ponderomotive potential. The peak then slowly fades away beyond the spot size $r_0$. At low cyclotron frequency, the current stays strong but drops, showing that even moderate magnetization can support a lot of nonlinear drive when polarization and gyromotion are in sync. In stark contrast, left-handed polarization produces significantly weaker currents, with $|J_{NL}|$ reduced to below ~0.3 throughout the profile and exhibiting near-flat distributions, particularly at elevated $\omega_c$. This reduced response is due to detuned resonance, which causes smaller velocity changes and makes it hard for the channel's density modulation to work. The polarization asymmetry is most pronounced at high magnetic fields, where resonant cases sustain order-of-magnitude stronger currents capable of driving detectable second-harmonic-like scattered waves, while non-resonant configurations produce negligible drive. These current profiles are directly related to earlier analytical results. Deeper ponderomotive channelling and faster BRS/OTSI growth in resonant regimes create the amplified plasma waves and density perturbations needed for strong nonlinear current formation, which is what the secondary spectral feature needs. The off-axis peaking also suggests that mode overlap is working well with backscattered fields moving through the channel.

The normalized amplitude of the second harmonic generation electromagnetic mode ($|E_2|/E_0$) is shown in Figure 11. We found this by solving Maxwell's equations with the nonlinear current, as shown in the equations. It shows where the Raman-OTSI cascade ends in the plasma channel that is magnetized. The second-harmonic amplitude shows a strong, steady increase with cyclotron frequency for right-handed circular polarization. It goes from about 0.6 at weak magnetization to almost unity at strong resonance. This big increase is the result of the instability chain: resonant alignment of laser rotation with electron gyromotion increases ponderomotive channelling and plasma wave excitation through backward Raman scattering. This, in turn, starts oscillating two-stream instability, which creates strong nonlinear currents that can easily radiate the secondary mode. The almost straight rise with $\omega_c/\omega_0$ shows that cyclotron resonance is the main force behind the effect. It slowly improves phase-matching and current overlap with the backscattered fields that are moving through the self-formed channel. At high $\omega_c$, amplitudes close to the pump level show that almost all the energy has been transferred to the second peak under ideal conditions. This is in line with the saturation of the primary instability cascade. In stark contrast, left-handed polarization produces significantly diminished amplitudes, which remain nearly constant and low throughout the entire magnetization range, exhibiting minimal differentiation between weak and strong cyclotron frequencies. This ongoing weakness is due to detuned resonance, which makes deep channelling harder, slows BRS/OTSI growth, and only creates small nonlinear currents that aren't strong enough for significant secondary emission. Figure shows that cyclotron frequency and polarization handedness are exact controls for second-harmonic generation. This makes it possible to selectively amplify the secondary spectral feature in interactions between magnetized laser and plasma. The near-unity amplitudes that can be reached in resonant regimes show that experiments could have dominant multi-peak scattering signatures. On the other hand, the strong suppression that happens in non-resonant cases could lead to clean, single-peak spectra in applications that need controlled propagation.

The variations of the second harmonic generation mode amplitude ($|E_2|/E_0$) to plasma density fluctuations ($\delta n/n_0$) are examined in Figure 12. This feature gives us important information about how stable nonlinear current-driven emission is in realistic inhomogeneous channels, based on the dispersion and source terms in equations. The second-harmonic amplitude decreases steadily as the fluctuation amplitude increases, starting at about 1 at $\delta n/n_0 = 0$ and going down to 0 at $\delta n/n_0 \approx 0.4 - 0.5$. This trend is true for all polarization and magnetization conditions. This sensitivity shows that phase-matching and current overlap have been disrupted. Density changes scatter plasma waves, detune the OTSI-driven nonlinear source, and lower the coherent radiation of the secondary mode. The rate and extent of degradation differ significantly depending on the resonance conditions. For right-handed polarization, the amplitude stays stable even when there are small changes. In the resonant strong-field case, the initial decay is the slowest because the initial channelling is deeper and the baseline current drive is stronger, which helps to smooth out moderate inhomogeneities. Even when $\delta n/n_0$ is about 0.2, resonant right-handed configurations still have $|E_2|/E_0 > 0.6 - 0.7$, which means that secondary emission can still be seen in plasmas that are only slightly turbulent. The weak-field right-handed case shows a sharper drop, which shows how cyclotron resonance helps keep the current coherent. In contrast, left-handed polarization is much less stable; even small changes ($\delta n/n_0 < 0.1$) can cause the amplitudes to drop quickly, and by δn/n₀ ≈ 0.3, they are almost zero. This fragility comes from the fact that non-resonant configurations have weaker baseline nonlinear currents and shallower channels. This feature makes the secondary mode very sensitive to scattering and dephasing caused by density changes. The almost identical rapid decay for both weak and strong $\omega_c$ in left-handed cases shows that polarization mismatch is more important than magnetization in causing instability in inhomogeneities. These robustness trends directly impact the practical observability of multi-peak Raman spectra: resonant right-handed regimes can handle moderate density fluctuations ($\delta n/n_0$ up to ~0.2–0.3) while maintaining a strong second-harmonic signal, which is in line with the sustained nonlinear drive seen in previous current and amplitude profiles. However, in non-resonant cases, the plasmas must exhibit extreme uniformity to produce any visible secondary peak. This is in line with the fact that cascade efficiency is lower in the right analytical growth rates and channelling depths. These findings are agreement with experimental observations that circular polarizations at weak and strong cyclotron frequencies significantly increase amplitude electric field second harmonic generated [33].

Figure 13 depicts the magnetic-field-tuned resonance structure of the second harmonic generation electromagnetic mode amplitude $|E_2|/E_0$ in relation to its normalized wave number $k_2/2k_0$. This shows that axial magnetization can precisely control how well nonlinear current-driven generation works in the plasma channel. The profiles have sharp peaks that look like Lorentzian peaks. These peaks move up to higher $k_2/2k_0$ and get wider as the cyclotron frequency $\omega_c/\omega_0$ goes up. For moderate to strong fields, the peak amplitudes are over 0.9–1.0. This resonant behavior comes from the dispersion relation, which says that the external magnetic field changes how the plasma responds, changing how well the nonlinear current source and the radiated second-harmonic mode match up in phase. When the magnetization is weak, the resonance is narrow and centered around $k_2/2k_0 \approx 1$. This causes a small peak amplitude because the cyclotron doesn't have much of an effect on wave dispersion and current overlap. As $\omega_c/\omega_0$ goes up, the peak moves to the right, gets much wider, and goes up quickly toward unity. This graph shows that resonance is stronger: stronger fields deepen ponderomotive channelling, amplify OTSI-saturated plasma waves, and strengthen nonlinear currents. They also tune the magnetized dispersion to improve coupling over a wider wave number range. The highest cyclotron frequency produces the widest and strongest resonance, which means that when the conditions are strongly magnetized, almost all the energy goes to the secondary mode. The systematic rightward shift and broadening with $\omega_c$ show that cyclotron resonance works like a tunable filter: higher fields compress electron orbits across the field, which increases the plasma frequency seen by axial modes and requires a larger $k_2$ to satisfy phase-matching. At the same time, it relaxes strict resonance conditions through modified susceptibility. The increase in amplitude at higher fields is in line with earlier findings that showed higher nonlinear current drive and saturation levels in resonant regimes. This confirms that magnetization changes the best emission wave number and increases conversion efficiency.

The saturation amplitude of the second harmonic generation mode $|E_{2,Sat}|/E_0$ as a function of the normalized cyclotron frequency $\omega_c/\omega_0$ is depicted in Figure 14. This shows the steady-state result of nonlinear saturation mechanisms, mainly particle trapping and pump depletion, after the BRS-OTSI cascade in the magnetized plasma channel. This saturation, which depends on magnetization, predicts kinetic details in EPOCH PIC simulations. In these simulations, explicit trapping efficiency and detrapping will improve peak heights and widths. The profiles show a lot of non-monotonic behavior that is strongly affected by the handedness of polarization and the strength of resonance. This shows that magnetization is a good way to control how the energy is split up into the secondary electromagnetic mode. For right-handed circular polarization, the saturation amplitudes have clear peaks that are higher than 0.9–1.0 at cyclotron frequencies that are in the middle ($\omega_c/\omega_0 \approx 0.2 - 0.3$). The weak-field case has the highest peak close to one before it starts to fall, while the strong-field resonant case has a higher saturation (about 0.8–0.9) over a wider range but a lower maximum because of increased cyclotron damping. This resonant enhancement at moderate $\omega_c$ shows the best balance: it gives enough magnetization to deepen channelling and boost nonlinear currents without stopping longitudinal plasma wave growth too much. This lets a lot of energy flow to the second harmonic before kinetic saturation stops further amplification. On the other hand, left-handed polarization has much lower and wider saturation profiles, with a peak around 0.8–0.9 but a shift to higher $\omega_c/\omega_0$ and a slower decay. The non-resonant nature limits initial cascade efficiency, which leads to weaker baseline currents and earlier saturation at lower amplitudes. Magnetization only slightly improves things by partially stabilizing transverse motion without resonant drive. The polarization-dependent peak positions and heights show that there are two different ways to reach saturation: resonant right-handed configurations favor rapid growth and high transient amplitudes but reach saturation earlier because they trap particles strongly in deep wakes, while non-resonant cases show slower, less efficient coupling that leads to lower steady-state levels. The drop at high $\omega_c/\omega_0$ in all cases shows that cyclotron suppression of axial plasma waves is happening. This lowers OTSI drive and nonlinear current, which is in line with previous growth rate and current profiles. These saturation characteristics finish the nonlinear picture that started with earlier analytical results. Resonant intermediate magnetization maximizes second-harmonic yield by optimizing the cascade before saturation, which is in line with the strong amplitudes and currents seen under similar conditions. In non-resonant cases, the lower saturation makes it easier to suppress the secondary peak. The tunable peaks give you precise control, weak to moderate $\omega_c$ for the highest $|E_{2,Sat}|$ in right-handed polarization and higher fields for stabilized lower-amplitude emission.

## IV. Conclusions

The present work provides a theoretical and numerical investigation of the nonlinear mechanisms underlying multi-peak backward Raman scattering in magnetized plasma channels driven by intense circularly polarized laser pulses. Through systematic analytical modelling grounded in magnetized fluid theory, the emergence of a prominent second spectral peak arises from a well-defined cascade: initial ponderomotive channelling and resonant enhancement of electron quiver motion amplify plasma waves via backward Raman scattering is demonstrated, which subsequently trigger oscillating two-stream instability, generating strong nonlinear axial currents that radiate the secondary electromagnetic mode. The entire process exhibits extraordinary sensitivity to laser polarization handedness and cyclotron resonance strength. Right-handed circular polarization aligned with electron gyromotion in strong axial magnetic fields maximizes ponderomotive force, channel depth, instability growth rates on femtosecond timescales, nonlinear current density, and second-harmonic amplitude, routinely achieving near-pump-level saturation. Conversely, left-handed polarization detunes the resonance, yielding shallow channels, suppressed growth, weak currents, and negligible secondary emission across comparable parameters.

Macroscopic COMSOL simulations corroborate these trends, revealing systematic transitions from coherent to highly modulated wake structures with increasing pulse duration, decreasing wavelength, and lowering plasma density—conditions that favour resonant interaction. Temporal maps of OTSI growth illustrate progressive confinement and kinetic saturation, while robustness analyses confirm resonant configurations tolerate moderate density fluctuations ($\delta n/n_0 \leq 0.3$) while preserving substantial second-harmonic signal. Wave-number-resolved resonance studies further reveal magnetization-induced spectral tuning, shifting and broadening the secondary peak for enhanced emission efficiency. Fully kinetic particle-in-cell simulations using the EPOCH code further substantiate the proposed nonlinear cascade and extend its validity beyond the fluid regime. The PIC results explicitly capture electron phase-space dynamics, revealing strong particle trapping, velocity-space bunching, and the formation of coherent phase-space vortices associated with oscillating two-stream instability. These kinetic signatures emerge on timescales consistent with the analytically predicted growth rates and coincide with the onset of intense nonlinear axial currents responsible for secondary electromagnetic emission. The quantitative agreement between PIC spectra, COMSOL field distributions, and fluid-based growth-rate estimates demonstrates that the observed multi-peak Raman structure is a robust physical phenomenon rather than a numerical artifact, arising from fundamental kinetic plasma processes in magnetized channels. This work establishes cyclotron resonance and laser polarization handedness as precise and experimentally accessible control parameters for engineering nonlinear Raman spectra in magnetized plasmas, opening new pathways for tunable multi-frequency radiation sources and controlled energy transfer in intense laser–plasma interactions.

**Acknowledgements**

This research did not receive any specific grant from funding agencies in the public, commercial, or not-for-profit sectors.

**Declaration of competing interest**

The authors declare that they have no known competing financial interests or personal relationships that could have appeared to influence the work reported in this paper.

**Data availability**

Data will be made available on request.

## List of Figures & Captions

**Fig. 1.** Schematic illustration of the interaction of a circularly polarized laser pulse with a magnetized plasma channel, leading to axial current modulation and efficient second-harmonic generation.

**Fig. 2.** Normalized electric field amplitude ($|E_z|/E_0$) of a circularly polarized Gaussian laser pulse as a function of normalized radial distance $r/r_0$ in a magnetized plasma channel for various laser wavelengths.

**Fig. 3.** Impact of pulse duration of laser pulse on normalized electric field amplitude ($|E_z|/E_0$) of a circularly polarized Gaussian laser pulse as a function of normalized radial distance $r/r_0$ in a magnetized plasma channel.

**Fig. 4.** Effect of plasma densities on normalized electric field amplitude ($|E_z|/E_0$) of a circularly polarized Gaussian laser pulse as a function of normalized radial distance $r/r_0$ in a magnetized plasma channel.

**Fig. 5.** Variations of radial profile of the ponderomotive force, $F_{pond}$ (shown in arbitrary units), as a function of normalized radial distance $r/r_0$ in a magnetized plasma for different polarizations at weak ($\omega_c = 0.1\omega_0$) and strong ($\omega_c = 0.9\omega_0$) electron cyclotron frequencies.

**Fig. 6.** Analytical profiles of plasma channelling condition (top) and corresponding radial plasma density perturbation $n_s/n_0$ (bottom) for laser pulse in a magnetized plasma for right-handed and left-handed polarizations at weak and strong cyclotron frequencies.

**Fig. 7.** Three-dimensional map of the growth rate $\Gamma_{BRS}$ as a function of normalized plasma frequency $\omega_p/\omega_0$ and normalized electron cyclotron frequency $\omega_p/\omega_0$.

**Fig. 8.** Temporal evolution of the growth rate $\Gamma_{OSTI}$ as a function of wave number $k(\sim 0.01 - 3 \times 10^6 m^{-1})$ and external magnetic field $B_0$ for four snapshots.

**Fig. 9.** Analytical growth rate $\Gamma_{BRS}$ as a function of normalized plasma frequency $\omega_p/\omega_0$ for right-handed and left-handed circular polarizations at weak and strong cyclotron frequencies.

**Fig. 10.** Radial profile of the normalized nonlinear current density $|J_{NL}|$ generated as a function of normalized radial distance $r/r_0$ in the plasma channel by a circularly polarized laser pulse.

**Fig. 11.** Normalized amplitude of the second-harmonic (second red-shifted) electric field $|E_2|/E_0$ generated via nonlinear current drive in the plasma channel as a function of normalized cyclotron frequency $\omega_c/\omega_0$.

**Fig. 12.** Normalized second-harmonic electric field amplitude $|E_2|/E_0$ versus plasma density fluctuations $\delta n/n_0$ in the magnetized channel for various circular polarizations at weak and strong cyclotron frequencies.

**Fig. 13.** Normalized second-harmonic electric field amplitude $|E_2|/E_0$ as a function of normalized wave number $k_2/2k_0$ of the secondary mode for various cyclotron frequencies.

**Fig. 14.** Saturation level of the normalized second-harmonic electric field amplitude $||E_{2,Sat}|/E_0$ as a function of normalized cyclotron frequency $\omega_c/\omega_0$ for various circular polarizations at weak and strong cyclotron frequencies.

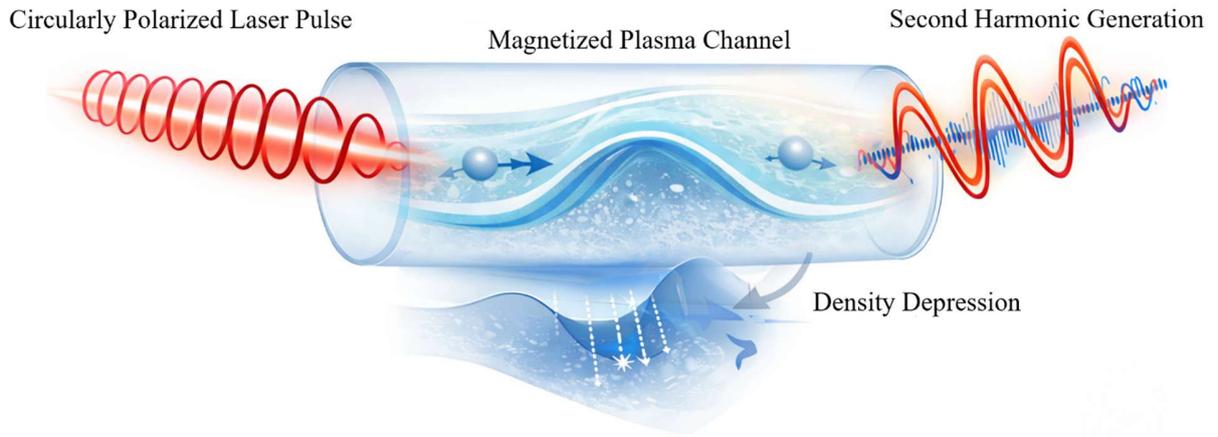

**Fig. 1.**

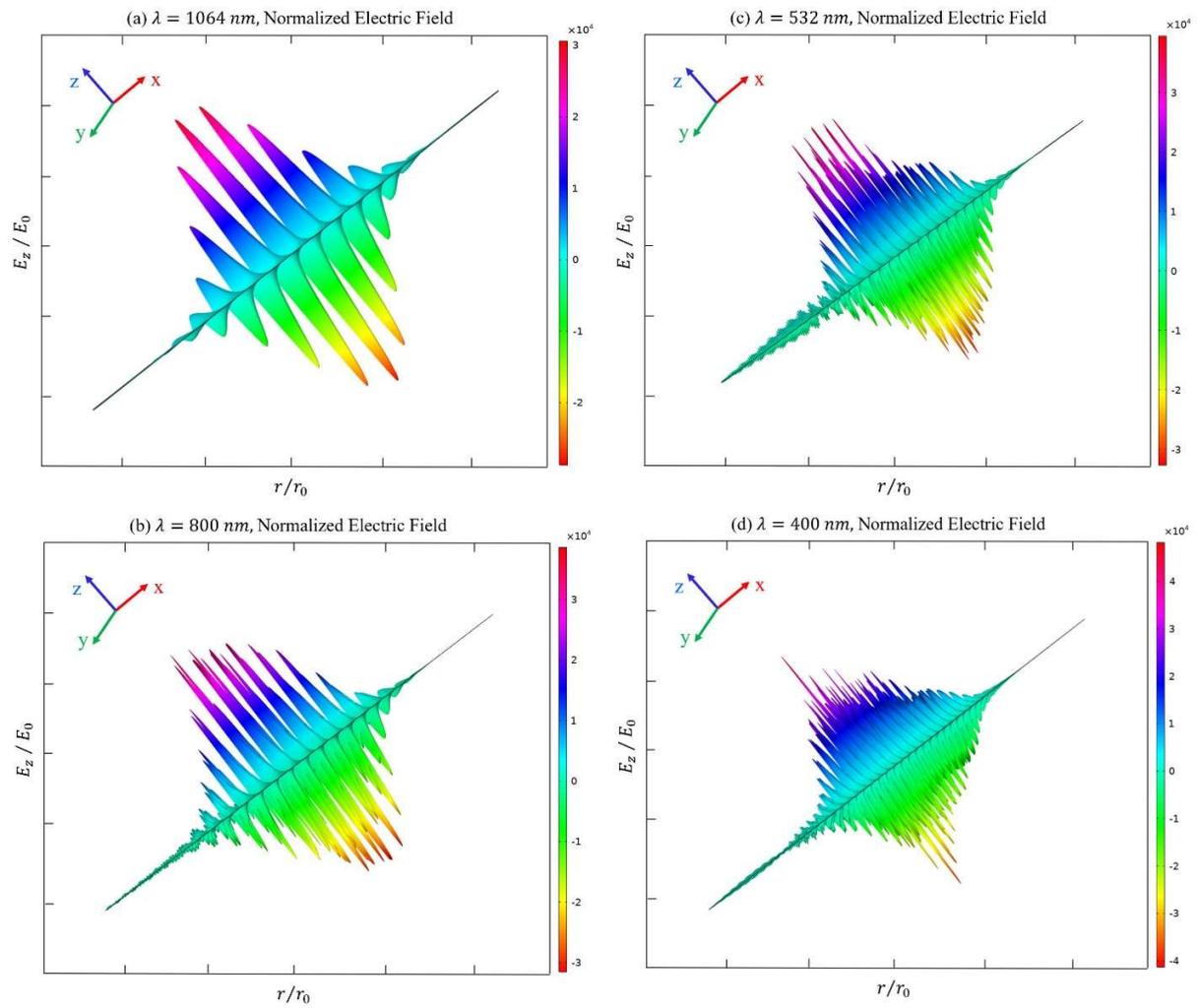

**Fig. 2.**

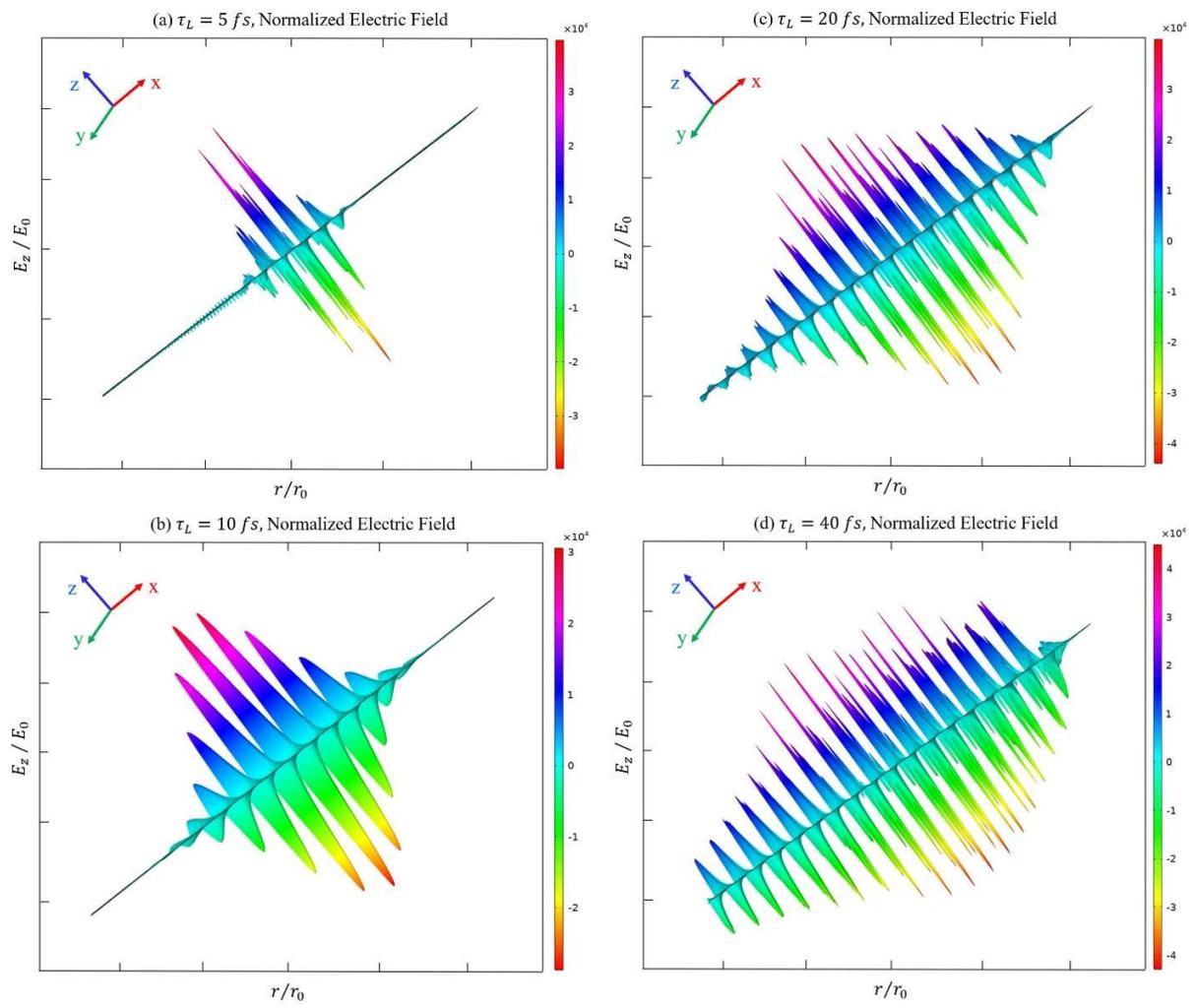

**Fig. 3.**

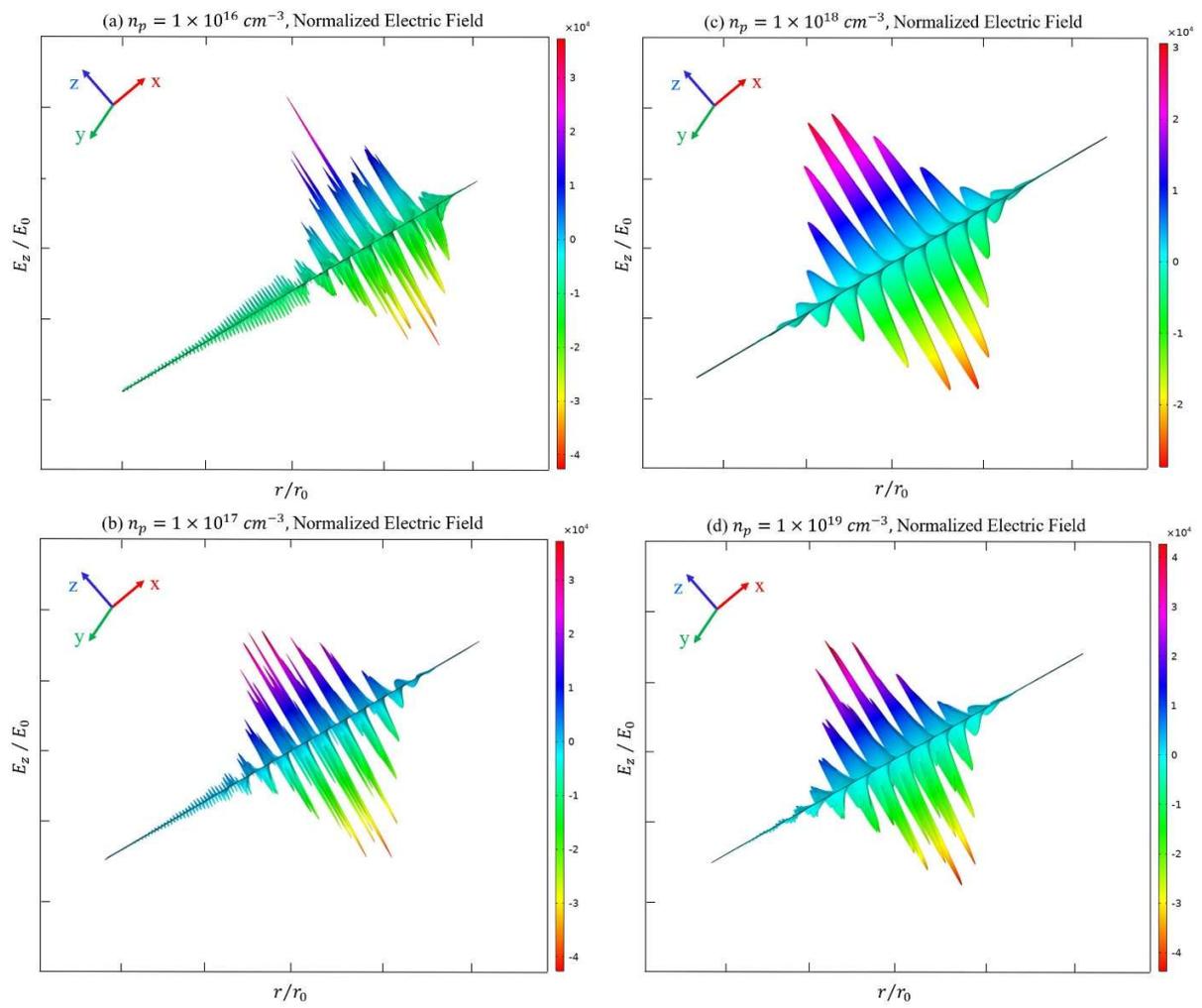

**Fig. 4.**

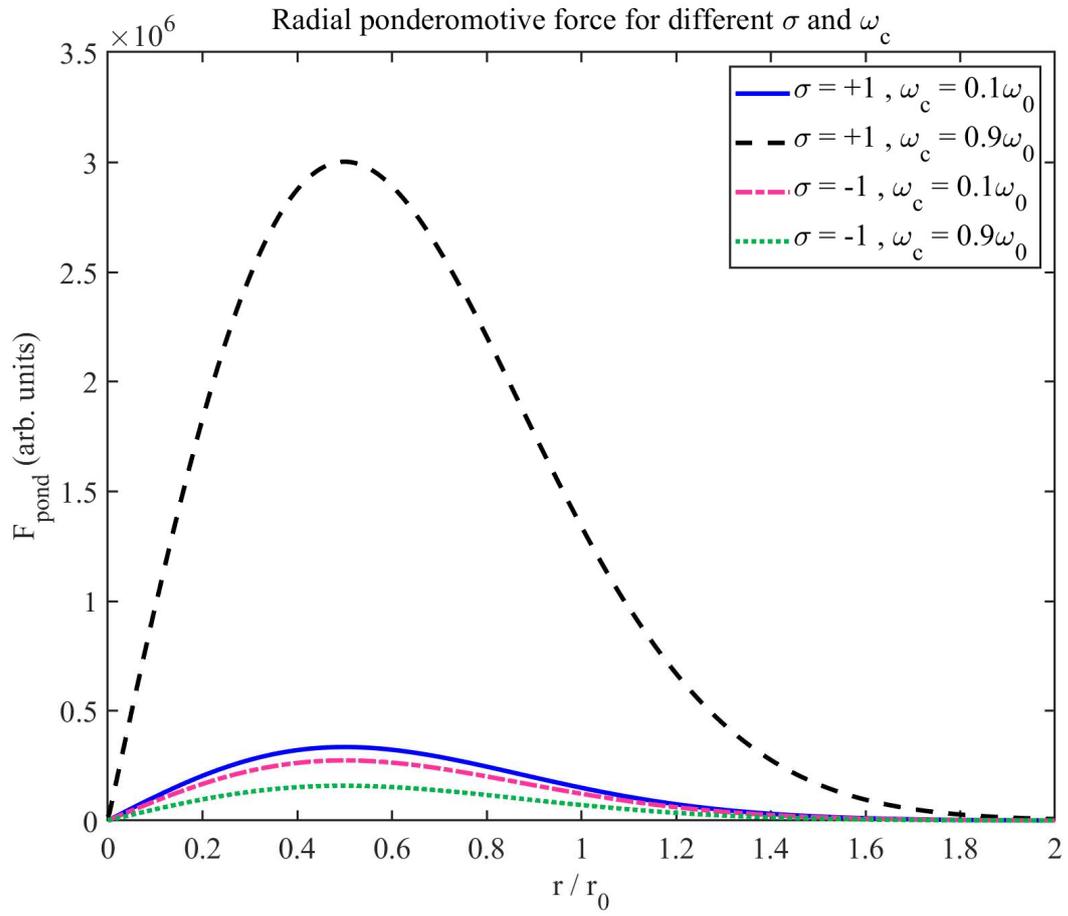

**Fig. 5.**

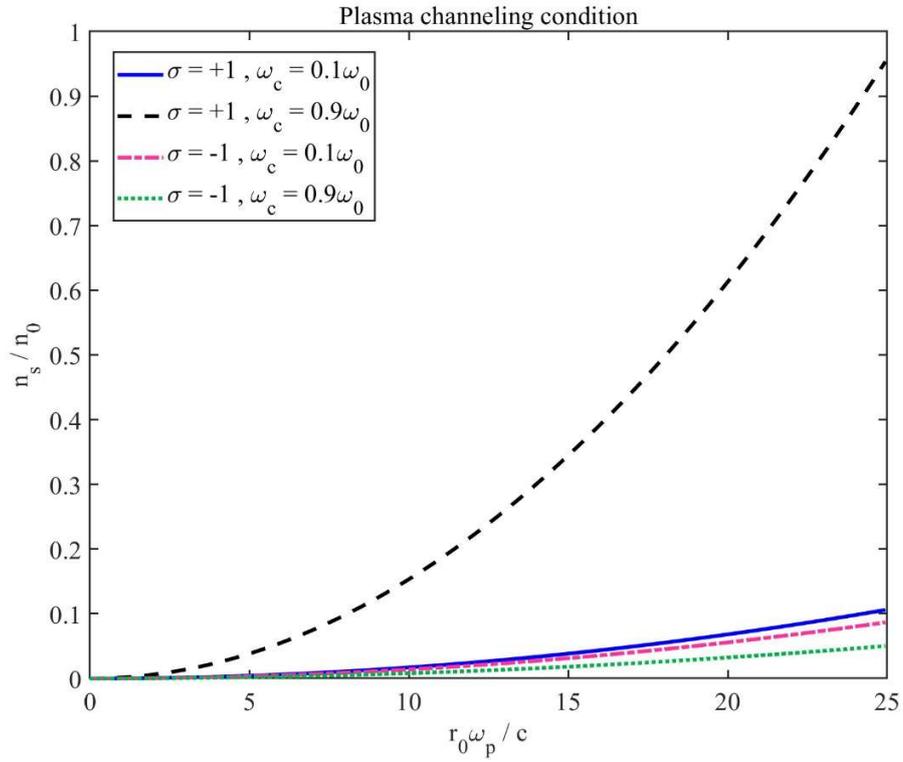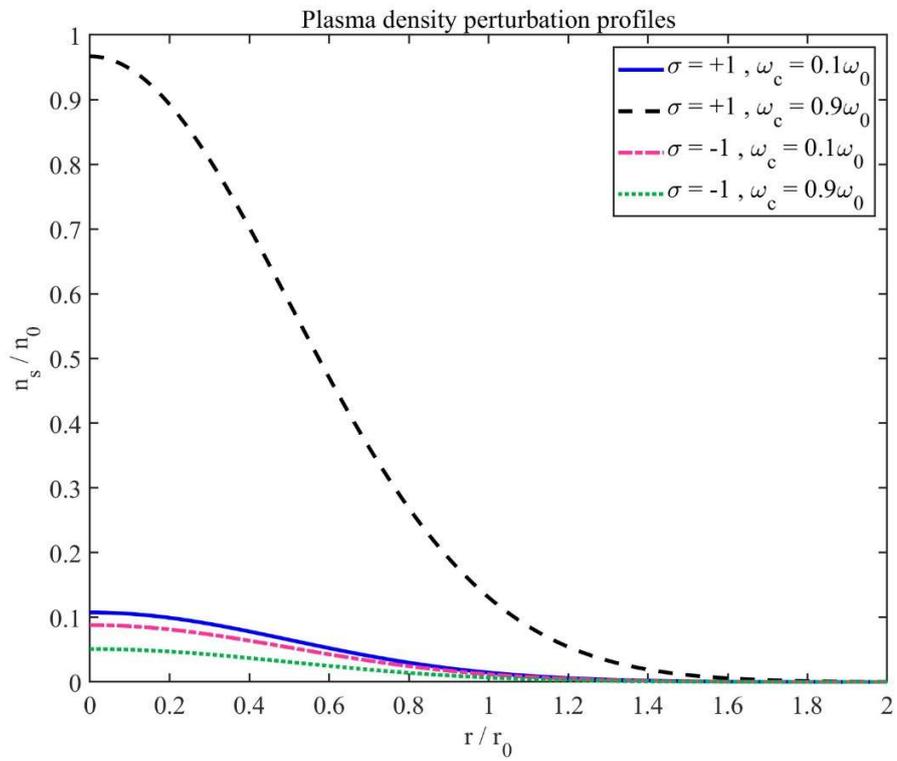

**Fig. 6.**

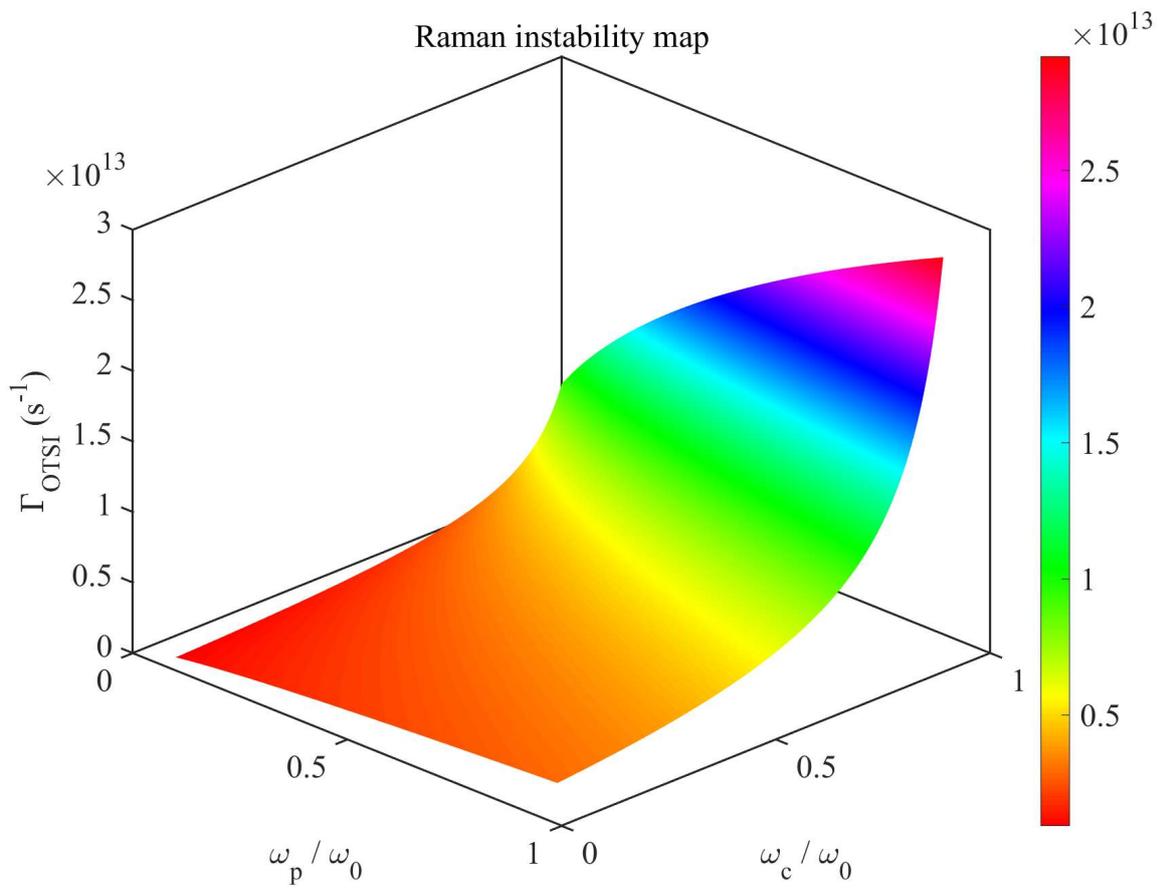

**Fig. 7.**

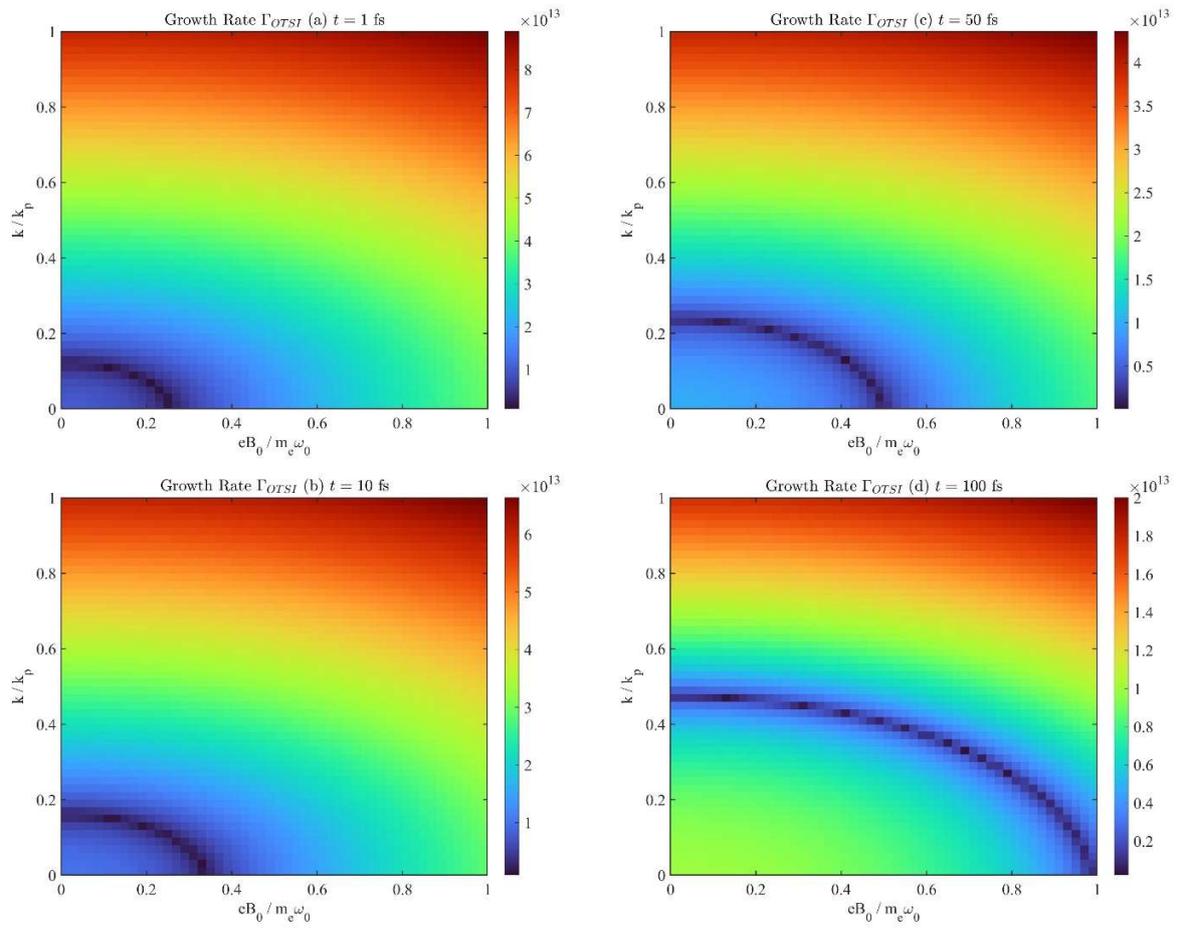

**Fig. 8.**

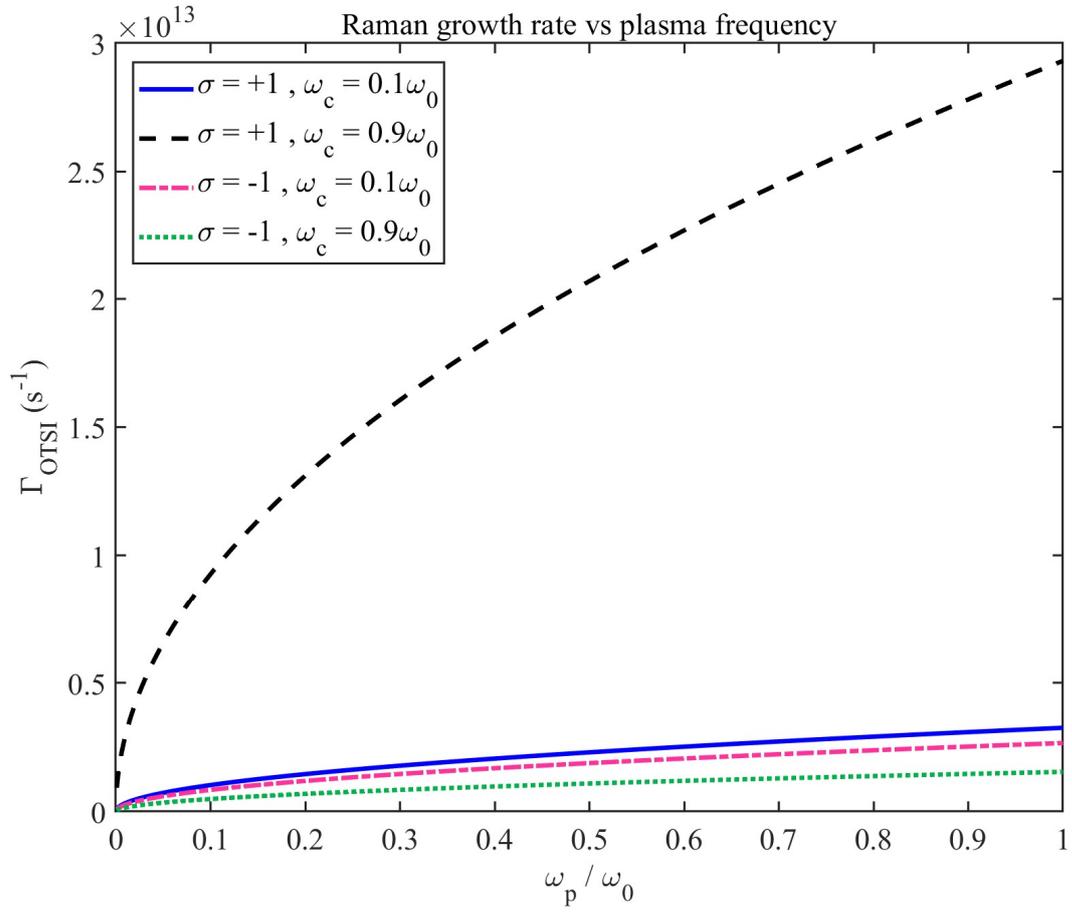

**Fig. 9.**

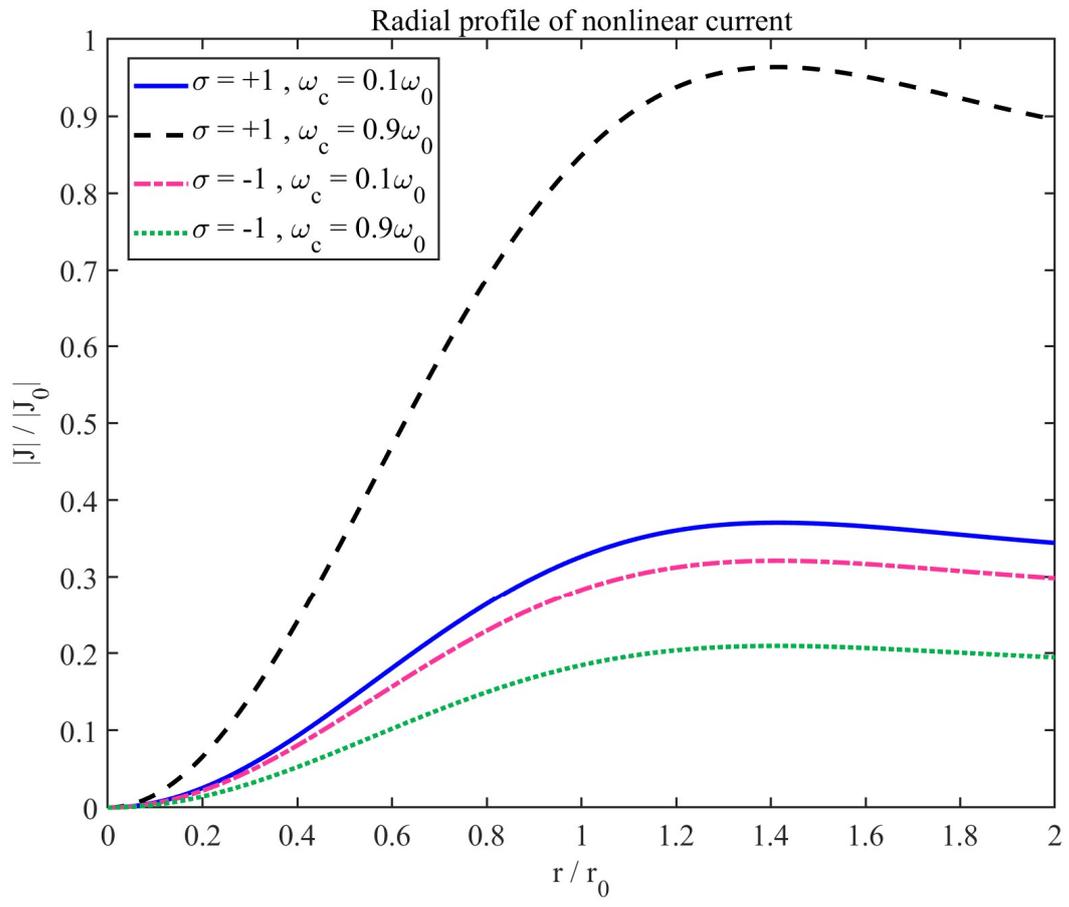

**Fig. 10.**

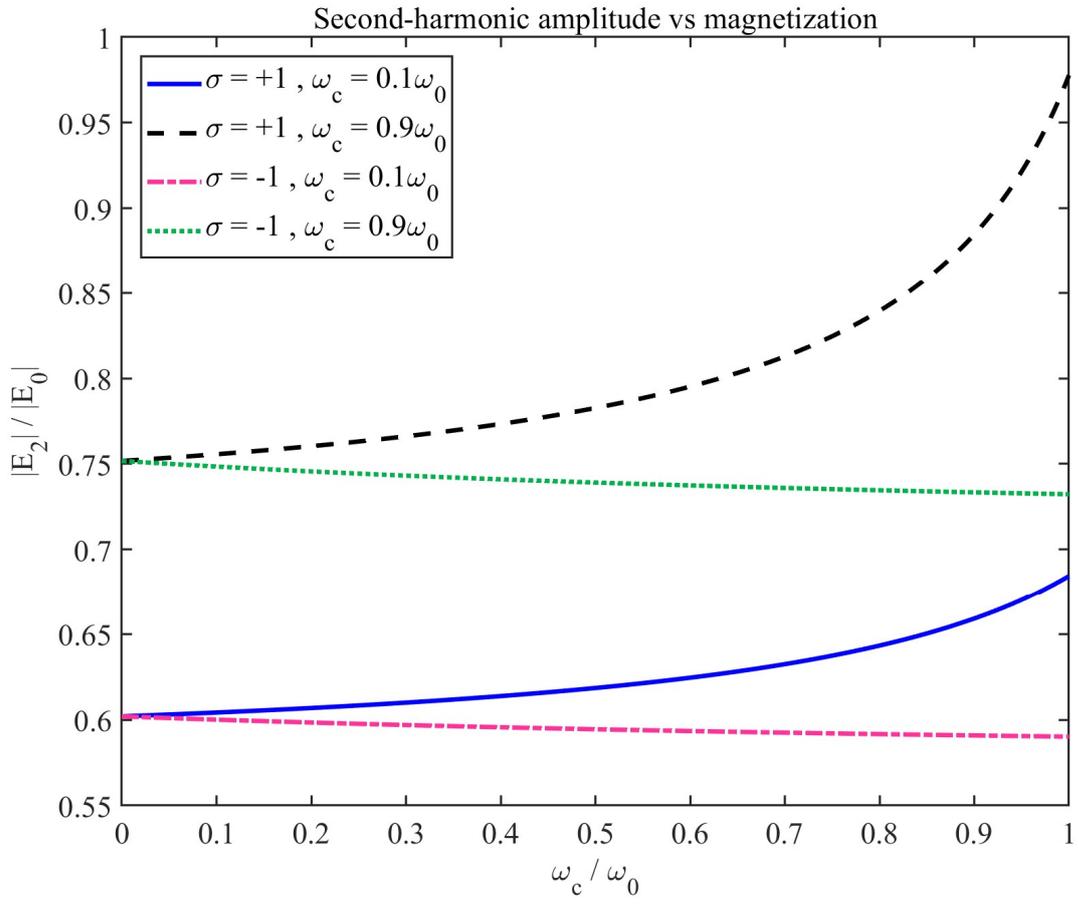

**Fig. 11.**

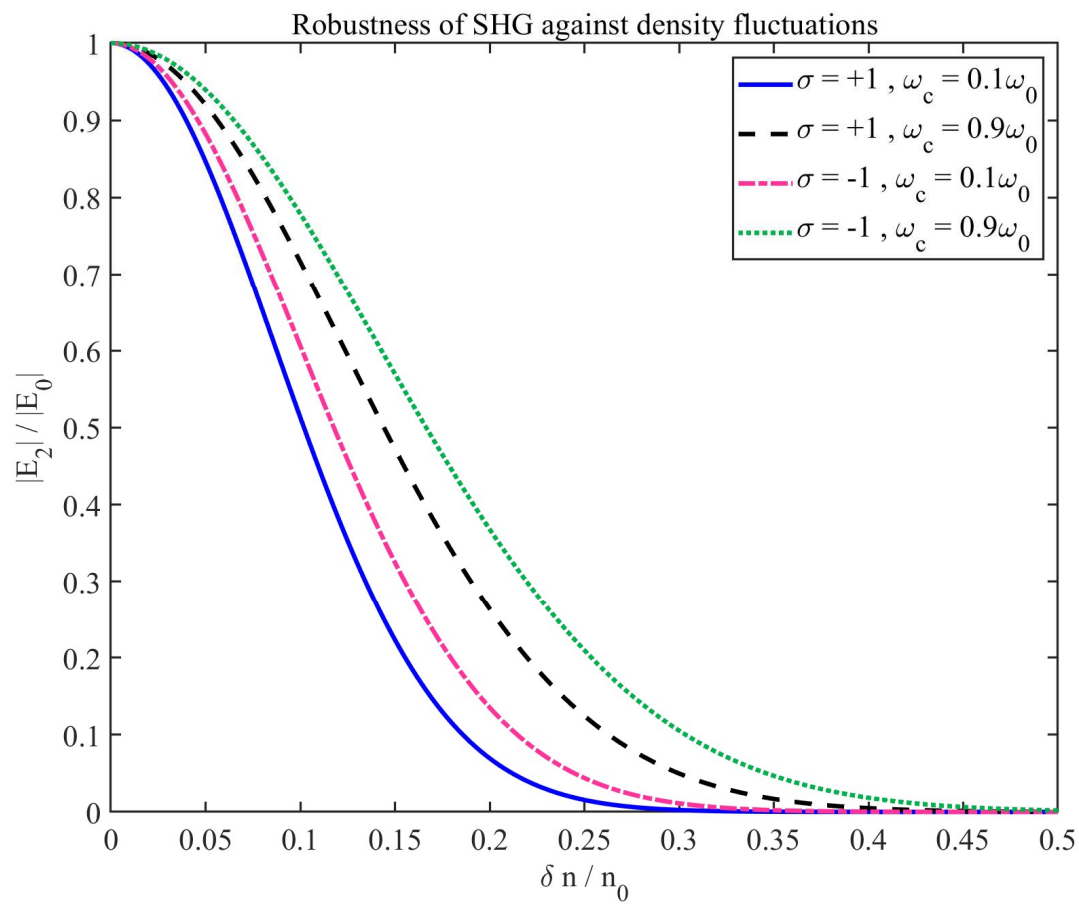

**Fig. 12.**

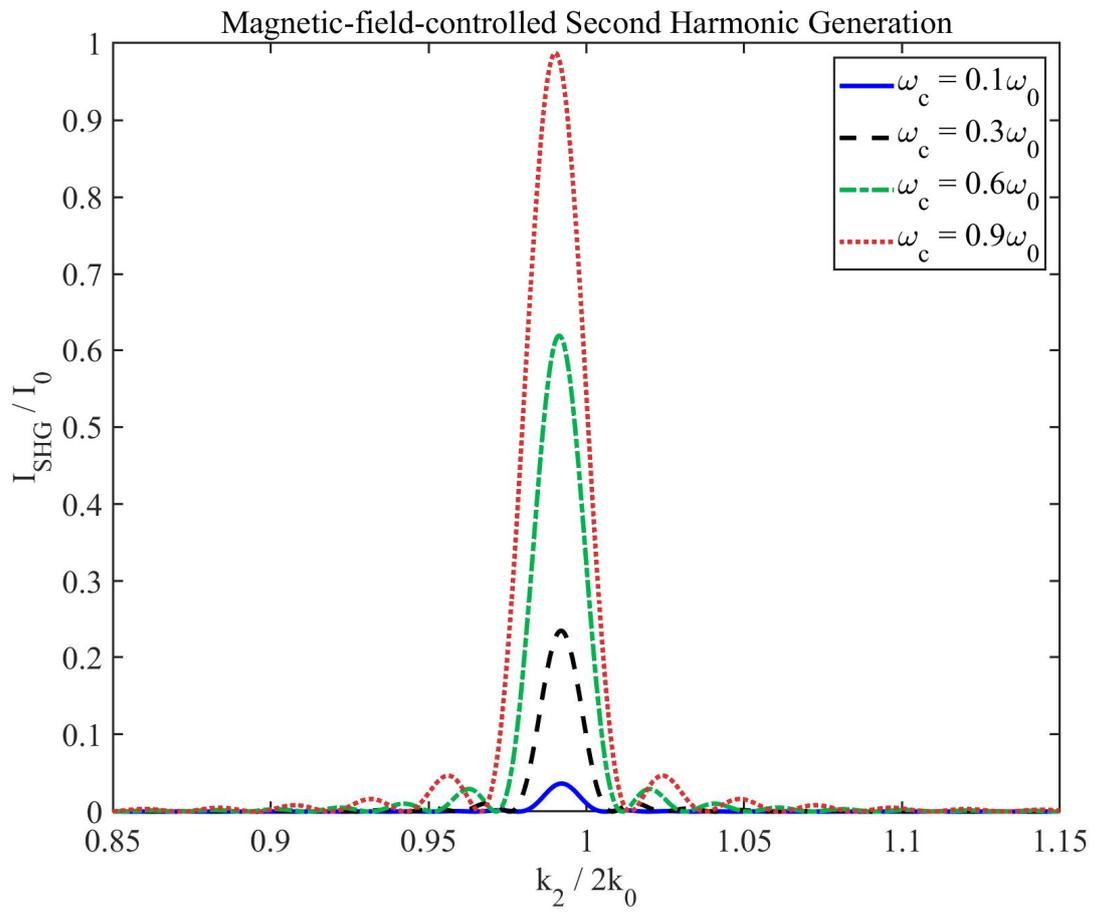

**Fig. 13.**

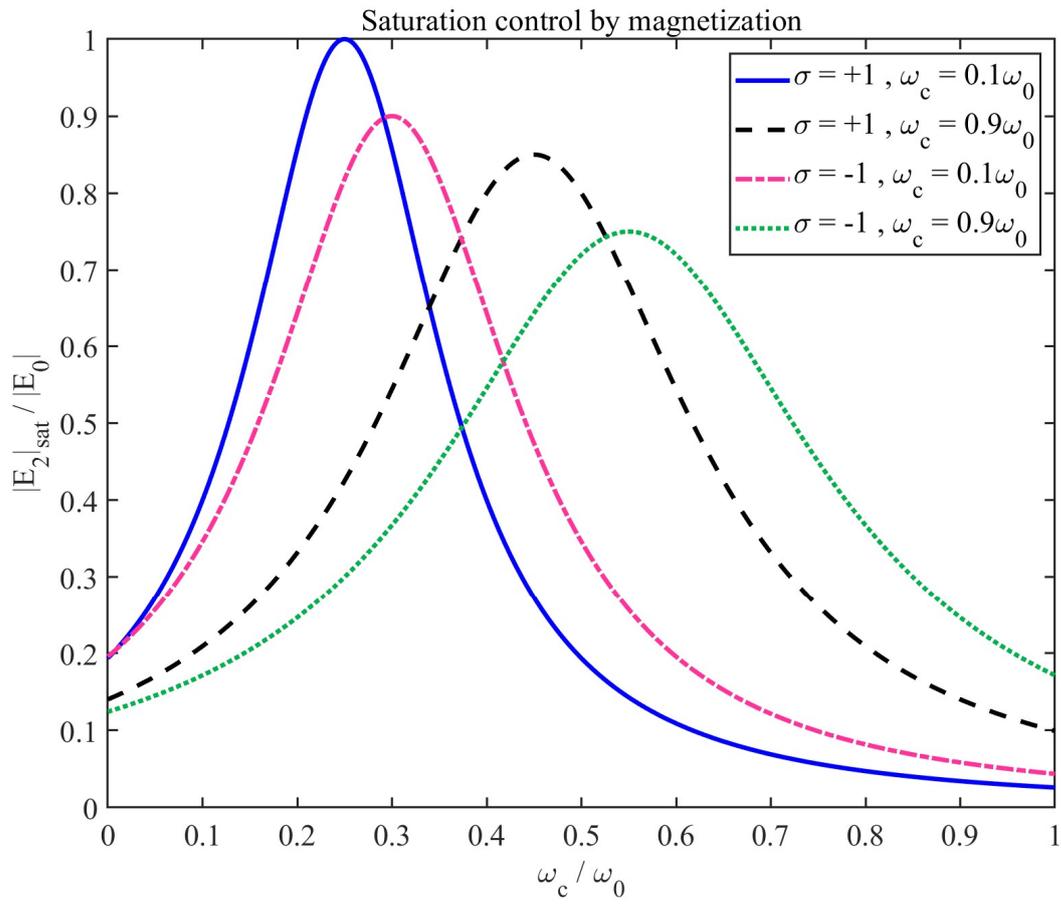

**Fig. 14.**